\documentclass{amsart}
\usepackage{graphicx}
\begin{document}

\title{Variations on Kak's Three Stage Quantum Cryptography Protocol}

\author{James Harold Thomas}
\address{Department of Electrical and Computer Engineering\\
Louisiana State University\\
Baton Rouge, Louisiana, 70803, USA} 

\begin{abstract}
This paper introduces a variation on Kak's three-stage quanutm key distribution protocol which allows for defence against the man in the middle attack. In addition, we introduce a new protocol, which also offers similar resiliance against such an attack.
\end{abstract}

\maketitle

\section{Introduction}
\label{intro}

In the relatively new field of quantum information science, there have been limited numbers of true breakthroughs. The quantum computer remains a machine on paper only, for a working realization has proved elusive. For one of the most powerful tools that the quantum computer would provide, namely the reduction of the factorization problem to polynomial time by Shor's algorithm, the most successful implementation to date has been to factor the number 15 into 3 times 5 [1].

The quantum computer might remain unrealized for years to come. Indeed, Kak has suggested that the current quantum circuit model for quantum computers is fundamentally flawed and new models must be developed to tackle the problem [2].

While there have been limited strides in some areas of quantum information science, one area in particular has produced realizable solutions in the field of cryptography. Quantum key distribution protocols have been successfully implemented and have produced commercially available products.

In this paper, we will discuss a new protocol proposed by Kak called the ``Three Stage Protocol.'' To Kak's protocol, we will introduce a modification which allows for greater security against man in the middle attacks.  In addition, we introduce a new single stage protocol which similarly allows for security against such attacks.

\section{Quantum Key Distribution}
The usefulness of quantum key distribution lies in the properties of the qubit, the quantum unit of information. Since a qubit is an object representing a quantum superposition state, the qubit cannot be copied. This is commonly called the no-cloning theorem [3]. This property ensures that during qubit data transmission, it is impossible for an evesedropper (Eve) to simply make copies of the qubits being sent, and thus manipulate these copied qubits to obtain the message. This useful property allows quantum data transmission to be used effectively in key distribution protocols as shown in [3].

When a private key can be transmitted securly along a quantum channel, then secure classical communication between the two parties (Alice and Bob) can be achieved using the private key (Figure 1).  One private key cryptosystem in use today is the Vernam cipher, also called the one time pad. According to Nielsen and Chuang, the security of the private key used in the Vernam cipher is sometimes ensured by transmitting it via such low-tech solutions as clandestine meetings or trusted couriers. The need for better transmission protocols is obvious.

\begin{figure}[htbp]
	\centering
		\includegraphics[width=0.75\textwidth]{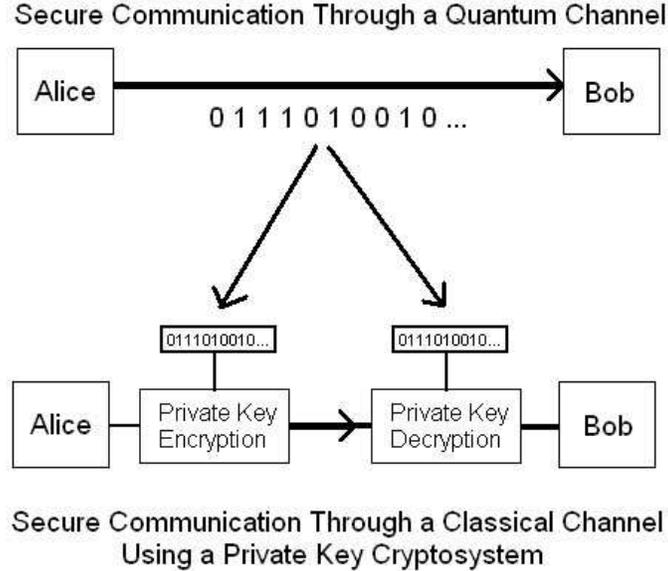}
	\caption{Quantum Key Distribution}
	\label{fig:figure_1}
\end{figure}

\section{Kak's Three-Stage Protocol}
In [4], S. Kak proposed a new quantum key distribution protocol based on secret unitary transformations (Figure 2). His protocol, like BB84, has three stages, but unlike BB84, it remains quantum across all three stages. In the first stage, Alice manipulates the message $X$, which is simply one of two orthogonal stages (e.g. $\alpha|0\rangle + \beta|1\rangle$ and $\beta|0\rangle - \alpha|1\rangle$) by means of a unitary transformation $U_{A}$, known only to her. Bob receives the new state, and in the second stage, applies his own secret transformation $U_{B}$, which is both a unitary transformation, and one that commutes with $U_{A}$, and sends the result back to Alice. In the third stage, Alice applies the Hermitian conjugate of her transformation, $U_{A}^{\dagger}$, and sends the result back to Bob. Since $U_{A}^{\dagger}U_{B}U_{A}(X) = U_{B}(X)$, Bob simply applies $U_{B}^{\dagger}$ and obtains the previously unknown state, $X$.

\begin{figure}[htbp]
	\centering
		\includegraphics[width=0.75\textwidth]{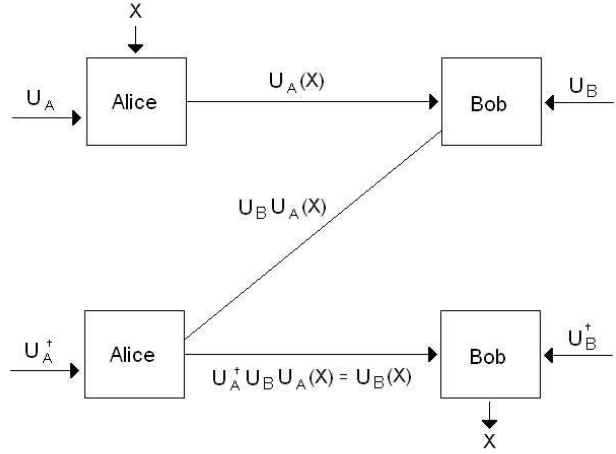}
	\caption{Kak's Three-Stage Protocol}
	\label{fig:figure_2}
\end{figure}

\section{The Man in the Middle Attack}
The Suceptability of both BB84 and Kak's three-stage protocol to man in the middle attacks has been documented [e.g. 5,6], and various methods to counter these attacks have been proposed [e.g. 7,8]. In such an attack, the eavesdropper, Eve, can attempt to thwart the communication between Alice and Bob in one of the following ways (Figure 3).

\begin{itemize}
\item[\bf{(1)}]	Eve receives the message from Alice by impersonating Bob. Eve then decodes Alice's message, and, now impersonating Alice, duplicates this message to Bob. In this scenerio, both Eve and Bob obtain the secret message.
\item[\bf{(2)}]	Eve impersonates Bob and decodes Alice's message as in scenerio 1, but instead of relaying the actual message to Bob, Eve relays a different message of her own choosing. In this scenerio, only Eve obtains the secret message.
\item[\bf{(3)}]	Eve impersonates Bob, but is not able to decode Alice's message. Instead, she impersonates Alice and sends her own message to Bob. In this scenerio, communication between Alice and Bob is blocked, but no secret message is comprimised.
\end{itemize}

\begin{figure}[htbp]
	\centering
		\includegraphics[width=0.75\textwidth]{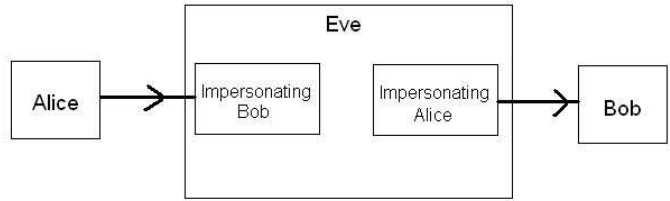}
	\caption{Man in the Middle Attack}
	\label{fig:figure_3}
\end{figure}

\section{A Variation on Kak's Three Stage Protocol}
In Kak's paper [4], he suggests using secret real valued orthogonal transformations to encrypt the qubits. Under orthogonal transformations of the same form (see below), the selection of the angles $\theta$ and $\phi$ by Alice and Bob respectively does not affect the outcome of the protocol. Furthermore, both Alice and Bob do not need to know what each other's angle selection is.

The reason that a man in the middle attack can be carried out is that when $U_{A}$ is assumed to be real valued, it is very easy for Eve to find another unitary transformation, $U_{E}$, which commutes with $U_{A}$. This is the underlying assumption by both Perkins [5], and Basuchowdhuri [7]. Indeed, for a 2x2 real valued unitary transformation (i.e. an orthogonal transformation), there is a limitation on its form.

Consider a 2x2 transformation:
\[
	\mathbf{U_{A}} = 
	\begin{bmatrix}
		w&x\\y&z
	\end{bmatrix}
,w,x,y,z \in \Re
\]
Then, for $U_{A}$ to be orthogonal (unitary), $U_{A}U_{A} = I$. This gives rise to the following equations:
\[
\begin{cases}
    w^2 + y^2 = 1 \\
    x^2 + z^2 = 1 \\
   	wx + yz = 0 
\end{cases}
\]
These equations are satisfied only when $U_{A}$ has one of the two following forms:
\[
	\mathbf{U_{1}(\theta)} =
	\begin{bmatrix}
		cos(\theta)&-sin(\theta)\\sin(\theta)&cos(\theta)
	\end{bmatrix},
\]
a rotation, as Kak proposed, or
\[
\\[10pt]
	\mathbf{U_{2}(\theta)} =
	\begin{bmatrix}
		cos(\theta)&sin(\theta)\\sin(\theta)&-cos(\theta)
	\end{bmatrix},
\]
a reflection across the line $\frac{\theta}{2}$.

The three-stage protocol demands that while Bob doesn't know the value of $\theta$ that Alice is using, he must know which of the two above forms of $U_{A}$ that Alice chooses. The reason for this is that while $U_{1}(\theta)$ commutes with $U_{1}(\phi)$ for any $\theta$ and $\phi$, and $U_{2}(\theta)$ commutes with $U_{2}(\phi)$ for any $\theta$ and $\phi$, $U_{1}(\theta)$ does not commute with $U_{2}(\phi)$ in general. So we will consider the choice of the form of $U_{A}$ to be public information. Once this information is known, Bob simply needs to choose his own angle $\phi$, and his transformation $U_{B}$ will be of the same form as $U_{A}$.
\\[10pt]
\[
U_{B} =
\begin{cases}
	U_{1}(\phi), U_{A} = U_{1}(\theta)\\
	U_{2}(\phi), U_{A} = U_{2}(\theta)
\end{cases}.
\]
\\[10pt]
It is easy to see that when the same form is used, $U_{A}$ and $U_{B}$ commute (i.e. $U_{A}U_{B} = U_{B}U_{A}$).
\\[10pt]
For Form 1:
\[
	U_{A}U_{B}=
	\begin{bmatrix}
		cos(\theta)&-sin(\theta)\\sin(\theta)&cos(\theta)
	\end{bmatrix}
	\begin{bmatrix}
		cos(\phi)&-sin(\phi)\\sin(\phi)&cos(\phi)
	\end{bmatrix}=
	\\[10pt]
\]
\[
	\begin{bmatrix}	cos(\theta)cos(\phi)-sin(\theta)sin(\phi)&-cos(\theta)sin(\phi)-cos(\phi)sin(\theta)\\cos(\phi)sin(\theta)+cos(\theta)sin(\phi)&-sin(\theta)sin(\phi)+cos(\theta)cos(\phi)
	\end{bmatrix}
\]
Applying trigonometric identities,
\[
U_{A}U_{B}=
	\begin{bmatrix}
		cos(\theta+\phi)&-sin(\theta+\phi)\\sin(\theta+\phi)&cos(\theta+\phi)
	\end{bmatrix}.
\]
Since $U_{A}$ and $U_{B}$ have the same form, it is clear that they commute.
\\[10pt]
For form 2:
\[
	U_{A}U_{B}=
	\begin{bmatrix}
		cos(\theta)&sin(\theta)\\sin(\theta)&-cos(\theta)
	\end{bmatrix}
	\begin{bmatrix}
		cos(\phi)&sin(\phi)\\sin(\phi)&-cos(\phi)
	\end{bmatrix}=
	\\[10pt]
\]
\[
	\begin{bmatrix}	cos(\theta)cos(\phi)+sin(\theta)sin(\phi)&cos(\theta)sin(\phi)-cos(\phi)sin(\theta)\\cos(\phi)sin(\theta)-cos(\theta)sin(\phi)&sin(\theta)sin(\phi)+cos(\theta)cos(\phi)
	\end{bmatrix}.
\]
Again, applying trigonometric identities,
\[
U_{A}U_{B}=
	\begin{bmatrix}
		cos(\theta+\phi)&sin(\phi-\theta)\\sin(\theta-\phi)&cos(\theta+\phi)
	\end{bmatrix}
\]

Since $U_{A}$ and $U_{B}$ always commute given the same form, then for Eve to impersonate Bob and obtain Alice's secret message, she only needs to select any angle $\psi$ and use it in her own transformation $U_{E}$ where $U_{E}$ is of the same form as $U_{A}$. Using $U_{E}$, Eve can obtain the secret state $X$ in the exact same way that Bob can obtain it. In addition, Eve can relay a message to Bob using her own $U_{E}$ transformation. Since Bob doesn't know what Alice's $U_{A}$ transformation is, Eve's $U_{E}$ is a valid substitution.

Suppose now instead of a orthogonal (i.e. unitary and real value) transformation, Alice chooses a more general complex valued unitary transformation,
\[
	\mathbf{U_{A}(\theta)} =
	\frac{1}{\sqrt{2}}
	\begin{bmatrix}
		e^{i\theta}&e^{-i\theta}\\ie^{i\theta}&-ie^{-i\theta}
	\end{bmatrix}
,\theta \in [0,2\pi).
\]
When Alice chooses a $U_{A}$ of this form, it is more difficult for Bob to find another transform, $U_{B}$ which commutes. We see that when 
\[
	\mathbf{U_{B}(\phi)} =
	\frac{1}{\sqrt{2}}
	\begin{bmatrix}
		e^{i\phi}&e^{-i\phi}\\ie^{i\phi}&-ie^{-i\phi}
	\end{bmatrix}
,\theta \in [0,2\pi),
\]
then
\[
	\mathbf{U_{A}U_{B}} =
	\begin{bmatrix}
		e^{i(\theta+\phi)}+ie^{i(\phi-\theta)}&e^{i(\theta-\phi)}-ie^{-i(\theta+\phi)}\\
		ie^{i(\theta+\phi)}+e^{i(\phi-\theta)}&ie^{i(\theta-\phi)}-e^{-i(\theta+\phi)}
	\end{bmatrix}
\]
and
\[
	\mathbf{U_{B}U_{A}} =
	\begin{bmatrix}
		e^{i(\phi+\theta)}+ie^{i(\theta-\phi)}&e^{i(\phi-\theta)}-ie^{-i(\phi+\theta)}\\
		ie^{i(\phi+\theta)}+e^{i(\theta-\phi)}&ie^{i(\phi-\theta)}-e^{-i(\phi+\theta)}
	\end{bmatrix}.
\]
These transformations commute only when $\phi = \theta + \pi$ (or any 2$\pi$ multiple). For Bob to decode Alice's message, he must have more information than simply the form of her transformation. He must know also the value of $\theta$ that she has chosen.

While this might seem to be a hindrence to the protocol, it allows for much greater security against a man in the middle attack. Eve attempts to intercept Alice's message to Bob by choosing a $U_{E}$ to impersonate Bob's $U_{B}$. As we saw earlier, when $U_{A}$ is real valued, Eve can simply pick any angle $\psi$ and generate a transformation $U_{E}$ that commutes. But with a complex valued $U_{A}$, Eve cannot guarantee a commuting transformation without knowing the value of $\theta$. Consider Eve's choice of a $\psi$ without knowledge of $\theta$. Then, $U_{A}^{\dagger}U_{E}U_{A} \ne U_{E}$.

\section{The Single Stage Quantum Cryptography Protocol}
When, as in the variation to Kak's protocol described above, Bob knows the value of $\theta$ that Alice has chosen for her transformation $U_{A}$ (assuming as above that the form of $U_{A}$ is public information), then he has full knowledge of $U_{A}$.  In this situation, Alice and Bob can forego the second two stages of the protocol and let Bob perform the transform $U_{A}^{\dagger}$ to obtain the unknown state $X$ (Figure 4). We have simply, $U_{A}^{\dagger}U_{A}(X) = X$. In this situation, there is no need for $U_{A}$ to be complex valued. We can have, as Kak proposed in [4],
\[
U_{A} =
	\begin{bmatrix}
		cos(\theta)&-sin(\theta)\\sin(\theta)&cos(\theta)
	\end{bmatrix}
\]
So for Eve to intercept the message $X$ and properly decode it, she would have to know the value of $\theta$.
The strength of this protocol is dependent on keeping the value of $\theta$ a secret known only to Alice and Bob. We enhance the security of our protocol by allowing for $\theta$ to change, which blocks any attempt by Eve at a statistical analysis of the qubits. We assume that before secure transmission may begin, there is some other secure protocol that Alice may use to transmit her initial value of $\theta$ to Bob. One example is Perkins' protocol which uses trusted certificates [6].

Suppose we restrict $\theta$ to the upper half plane of the unit circle. After $l$ qubits are successfully transmitted from Alice to Bob, the qubits $l+1$ to $l+k$ will be used to obtain the new value of $\theta$. The $k$ data bits selected by Alice for these $k$ qubits will represent an integer $N$ such that if $b_{n}$ is the $n$th bit transmitted ($b_{n} \in \{0,1\}$), then $N=b_{l+1}+2b_{l+2}+4b_{l+3}+...+2^{k}b_{l+k}$. When these four qubits are received by Bob and decoded, Alice and Bob adjust their transformations $U_{A}$ and $U_{A}^{\dagger}$ respectively such that
\[
U_{A} =
	\begin{bmatrix}
		cos(\theta_{N})&-sin(\theta_{N})\\sin(\theta_{N})&cos(\theta_{N})
	\end{bmatrix}
	, \theta_{N}=\frac{N\pi}{2^{k}}.
\]
After this, Alice transmits $l$ more qubits to Bob before again changing the value of $\theta$. In this fashion, any attempt by Eve to obtain the value of $\theta$ with no prior knowledge would be extremely difficult.


\begin{figure}[htbp]
	\centering
		\includegraphics[width=0.75\textwidth]{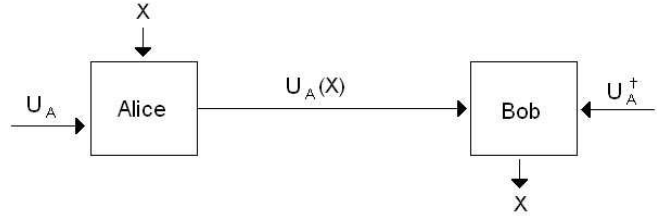}
	\caption{The Single Stage Quantum Cryptography Protocol}
	\label{fig:figure_4}
\end{figure}

\begin{figure}[htbp]
	\centering
		\includegraphics[width=0.75\textwidth]{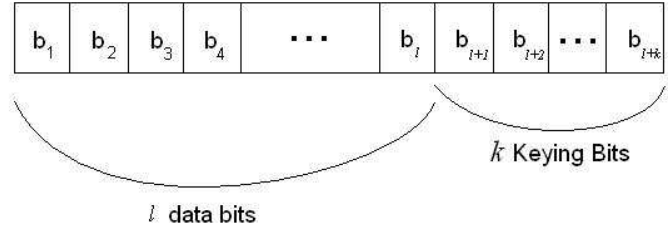}
	\caption{Framing on the Single Stage Protocol}
	\label{fig:figure_5}
\end{figure}

\section*{Acknowledgements}
The author thanks the Louisiana Board of Regents, BoRSF, under agreement NASA/LEQSF(2005-2010)-LaSPACE and NASA/LaSPACE under grant NNG05GH22H for support during this project.

\end{document}